\pdfoutput=1
\documentclass[10pt,conference]{IEEEtran}
\usepackage[T1]{fontenc}
\usepackage[utf8]{inputenc}

\usepackage{cite}

\usepackage{amsmath}
\usepackage{amssymb}

\usepackage{array}
\usepackage{booktabs}

\usepackage{graphicx}
\usepackage[dvipsnames]{xcolor}

\usepackage[hidelinks]{hyperref}
\usepackage{url}

\usepackage[capitalise,nameinlink]{cleveref}

\usepackage{tikz}
\usetikzlibrary{arrows.meta,positioning,calc}
\definecolor{wfblue}{HTML}{0F4D92}
\definecolor{wfbluebg}{HTML}{EAF1F9}
\definecolor{wfgreen}{HTML}{3C8D3C}
\definecolor{wfred}{HTML}{B64342}
\definecolor{wfgray}{HTML}{767676}
\definecolor{wfpanel}{HTML}{F2F4F7}
\definecolor{wfline}{HTML}{D9DEE5}
\usepackage{fontawesome5}

\usepackage[most]{tcolorbox}
\newtcolorbox{finding}[1]{enhanced, breakable, colback=wfbluebg, colframe=wfblue,
  boxrule=0.5pt, leftrule=2.2pt, arc=1pt, left=4pt, right=4pt, top=2pt, bottom=2pt,
  before skip=5pt, after skip=5pt, fontupper=\small,
  title={#1}, fonttitle=\small\bfseries\sffamily, coltitle=wfblue, colbacktitle=wfbluebg,
  toptitle=1.5pt, bottomtitle=0pt, titlerule=0pt}

\newcommand{\para}[1]{\par\smallskip\noindent\textbf{#1}\enspace}

\usepackage{fancyvrb}
\definecolor{snipnavy}{HTML}{0F4D92}\definecolor{snipblue}{HTML}{3775BA}\definecolor{snipred}{HTML}{B64342}
\definecolor{snipgrey}{HTML}{8A93A0}\definecolor{snippanel}{HTML}{F2F4F7}\definecolor{snipedge}{HTML}{CDD3DC}
\definecolor{sniphlC}{HTML}{DCE8F6}\definecolor{sniphlR}{HTML}{F8DCDA}
\colorlet{Ccolor}{snipblue}\colorlet{Rustcolor}{snipred}
\newcommand{\sniptt}{\fontfamily{zi4}\selectfont}
\newtcolorbox{snipcard}[3]{enhanced, colback=white, colframe=snipedge, boxrule=0.4pt, arc=2pt,
  left=3pt, right=3pt, top=1pt, bottom=2pt, before skip=0pt, after skip=0pt,
  title={\textbf{#1}\ \textperiodcentered\ {\sniptt #2}\hfill
         \href{#3}{\color{snipblue}\scriptsize\mdseries source\,$\nearrow$}},
  fonttitle=\sffamily\footnotesize, colbacktitle=snippanel, coltitle=snipnavy, titlerule=0pt,
  toptitle=2pt, bottomtitle=0pt}
\newcommand{\snipnote}[1]{{\sffamily\scriptsize\itshape\color{black!72}#1\par}\vspace{2pt}}
\newcommand{\sniplabel}[3]{\par\vspace{2pt}{\sffamily\scriptsize\textbf{\color{#1color}#1}}\hspace{0.5em}%
  \href{#3}{\sniptt\scriptsize\color{snipblue}#2\,$\nearrow$}\par\vspace{1pt}}
\tcbset{snipcodebox/.style={enhanced, colback=white, colframe=snipedge, boxrule=0.3pt, arc=0pt,
  left=0pt, right=1pt, top=0.5pt, bottom=0.5pt, before skip=0pt, after skip=3pt}}
\newcommand{\snipfont}{\sniptt\fontsize{6.1pt}{7.3pt}\selectfont}

\DefineVerbatimEnvironment{snipcodeC}{Verbatim}{commandchars=\?\@\", formatcom=\snipfont,
  frame=leftline, framerule=1.6pt, rulecolor=\color{snipblue}, framesep=1pt, xleftmargin=0pt}
\DefineVerbatimEnvironment{snipcodeR}{Verbatim}{commandchars=\?\@\", formatcom=\snipfont,
  frame=leftline, framerule=1.6pt, rulecolor=\color{snipred}, framesep=1pt, xleftmargin=0pt}
\definecolor{snipkw}{HTML}{7A3E9D}\definecolor{snipty}{HTML}{0E7490}\definecolor{snipnumc}{HTML}{B35C00}
\definecolor{snipstr}{HTML}{2E7D32}\definecolor{snipcom}{HTML}{8A93A0}\definecolor{snipfn}{HTML}{1F4E99}

\title{Re-Testing LLM-Based C-to-Rust Translations\\ Against Their Original C}

\author{\IEEEauthorblockN{Muhammad Hassnain}
\IEEEauthorblockA{University of California, Davis\\
\texttt{mhassnain@ucdavis.edu}}}
\hypersetup{pdftitle={Re-Testing LLM-Based C-to-Rust Translations Against Their Original C},
  pdfauthor={Muhammad Hassnain}}

\begin{document}
\pagestyle{plain} 

\maketitle
\thispagestyle{plain}

\begin{abstract}
Most C-to-Rust translators use large language models (LLMs), and each reports a success rate from its own correctness check. We survey $34$ C-to-Rust translators that use LLMs. Only one reports testing its check, and eleven release the original C and their Rust in runnable form. We re-test all eleven. We run the C and the Rust on generated inputs, compare every output, and count a divergence only where the C has no undefined behavior. All eleven
release defective translations: we find $160$ defects, $93$ of them outright errors on ordinary
inputs. Their own checks accepted $86$ defective translations. For $81$ of them, the released check never compared the programs on an input that exposes the defect. For three of the other five, it ran one but compared only part of the output. A success rate is only as good as its check. Before counting a translation as a success, authors should compare every output with the C's on inputs the repair loop never saw, check that a verifier reaches its assertion, release exactly what the check accepted, and report how they tested the check.
\end{abstract}


\section{Introduction}
\label{sec:intro}

Memory-safety errors are a leading cause of security vulnerabilities in C and C++ software, and the
U.S.\ government has urged a move to memory-safe languages such as Rust~\cite{oncd}. Rewriting
existing C by hand is costly, so DARPA's TRACTOR program~\cite{tractor} and a growing number of
research systems translate C to Rust automatically, most with large language models (LLMs). An LLM can produce Rust that compiles yet behaves differently from the C. Each system therefore checks its translations with a fuzzer, a model checker, symbolic execution or tests, and reports the share that pass as its success rate.

Memory safety alone does not make code trustworthy, even in safe Rust~\cite{counterexamples-safe-rust};
trust needs evidence, such as test and verification results~\cite{cargo-sherlock}. A success rate is only as good as the check behind it, yet few papers examine their checks.
We survey $34$ C-to-Rust systems (\cref{sec:survey}). Of the $25$ whose repair loop runs inputs, six report results on inputs the loop never used. Only one tests its check on known-wrong translations. Eleven release the
original C and their own Rust in runnable form.

\begin{figure*}[t]
\centering
\includegraphics[width=\textwidth]{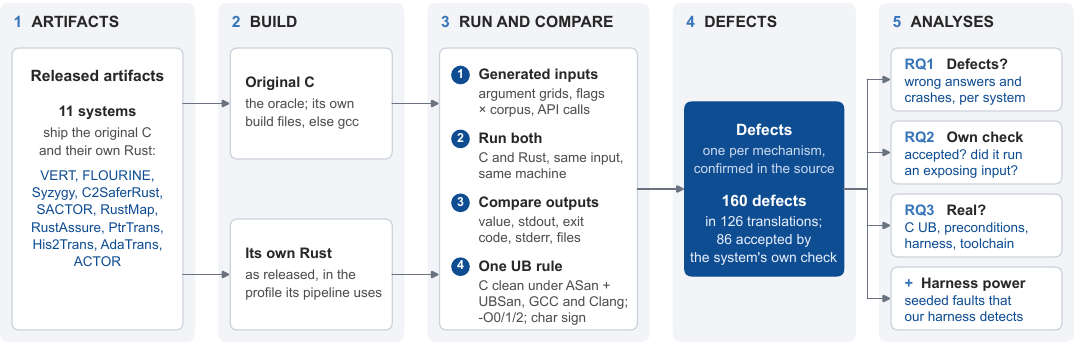}
\caption{Re-testing released translations. For every system whose artifact ships the original C
and its own Rust (1), we build both (2), run them on generated inputs, compare every output, and
keep a divergence only on inputs where the C passes our UB rule (3). We group the divergences into defects, one per mechanism, and confirm each in the translated source (4). The analyses (5) answer
RQ1-RQ3 and measure our harnesses' power by planting bugs.}
\label{fig:workflow}
\end{figure*}

We re-test the released translations of all eleven (\cref{fig:workflow}). The original C is
the \emph{oracle}: the translation claims to preserve its behavior. For each system, a test
\emph{harness} runs the C and the Rust on generated inputs and compares every output; a divergence counts only on inputs where we find no undefined behavior (UB) in the C. Reading the Rust source, we
group divergences into \emph{defects}, one per faulty mechanism. From each artifact we then measure
whether the system's own check \emph{accepted} each defective translation and ran an exposing input. We ask:
\begin{description}
  \item[RQ1:] Do released translations preserve the behavior of their original C on inputs where the
  C is well-defined?
  \item[RQ2:] What did each system's own check record for the translations that do not, and did it
  run an input that exposes the difference?
  \item[RQ3:] How robust are these findings to undefined behavior and preconditions in the C, to the
  toolchain and build profile, and to the power of our harnesses?
\end{description}

All eleven release defective translations (RQ1): $160$ defects, of which $93$ are \emph{outright errors}, wrong outputs or crashes on ordinary inputs. Their checks accepted $86$ defective translations, $56$ with an outright error (RQ2). For $81$ of the $86$, the released check never compared the programs on an exposing input. For three of the other five, it ran one but compared only part of the output. FLOURINE's own check rejected $20$ files that its artifact ships as successes; six are defective. Mutation analysis measures our harnesses' power (RQ3): they catch from $20\%$ of planted bugs (C2SaferRust) to $91\%$ (PtrTrans).

We contribute a survey of $34$ systems (\cref{sec:survey}), a method for re-testing released translations
against their original C (\cref{sec:method}), and a re-test of all eleven with measured harness power
(\cref{sec:results}).

\section{Selection of Translators}
\label{sec:survey}

We include every C-to-Rust system that uses an LLM to write or repair the Rust and was first public
by 31 October 2026. Its paper must contribute a translation pipeline, not a benchmark or study,
and report a success rate from an automated correctness check. We searched arXiv and Crossref and
followed the related work of each system we found. Of $69$ C-to-Rust candidates, $34$ qualify. From each paper we record whether the evaluation is \emph{held out}, using inputs that the repair loop (where the LLM revises the Rust) never saw. We also record whether the authors \emph{mutation-test} the check by feeding it known-wrong translations.

Six of the $34$ report a held-out result~\cite{actor,reboot,adatrans,actoruw,orbit,pr2}, two of them
for a subset only. Nineteen do not. In eight, the repair loop runs no inputs, so every evaluation
input is new to it. Heimdall's check is symbolic and runs no inputs at all; it is the one system that
mutation-tests its check~\cite{heimdall}. No system does both.

We re-test every artifact that ships the original C and the system's own Rust, both runnable in
user space. Eleven systems qualify (\cref{tab:survey}). Heimdall's programs are eBPF kernel hooks, not user-space programs, and we could not build the C that Rustine~\cite{rustine} ships.

\begin{table}[t]
\centering
\caption{The surveyed systems whose artifacts meet our re-test rule (all: every shipped translation
qualifies; part: some do). Held-out: evaluated on inputs its repair loop never used (-: the loop
runs no inputs). Mut.: check mutation-tested.}
\label{tab:survey}
\footnotesize
\setlength{\tabcolsep}{3pt}
\begin{tabular}{@{}l>{\raggedright\arraybackslash}p{32mm}cc>{\centering\arraybackslash}p{6mm}@{}}
\toprule
System & Own check & Held-out & Mut. & Art. \\
\midrule
VERT~\cite{vert}               & fuzzer, model checker vs.\ a Rust copy of the C & No$^\ddagger$ & No & all \\
FLOURINE~\cite{flourine}       & cross-language fuzzing        & No & No & all \\
Syzygy~\cite{syzygy}           & equivalence tests vs.\ the C  & No & No & all \\
C2SaferRust~\cite{c2saferrust} & coreutils end-to-end tests    & No & No & all \\
SACTOR~\cite{sactor}           & end-to-end tests via FFI      & No & No & part \\
RustMap~\cite{rustmap}         & functional tests; log diff    & No & No & all \\
RustAssure~\cite{rustassure}   & KLEE symbolic testing         & - & No & part \\
PtrTrans~\cite{ptrtrans}       & equivalence unit tests        & - & No & part \\
His2Trans~\cite{his2trans}     & project tests                 & - & No & part \\
AdaTrans~\cite{adatrans}       & fuzzing oracle                & Yes$^\S$ & No & all \\
ACTOR~\cite{actoruw}           & benchmark hidden tests        & Yes & No & all \\
\bottomrule
\end{tabular}

\vspace{2pt}
\parbox{\columnwidth}{\scriptsize $^\ddagger$For the fuzzer; its model checker runs no inputs. $^\S$Per its
Sec.~5.1; its Sec.~3.4 says the repair loop also runs a fuzzing suite.}
\end{table}

\section{Method}
\label{sec:method}

\para{What we count.}
A system releases \emph{translations}: functions, libraries or programs. One run of the C and the
Rust on the same input is a \emph{case}; the case \emph{diverges} if their outputs differ. For each
divergence we find the cause in the Rust source. A \emph{defect} is one such cause: a mechanism that makes the translation give a wrong output or crash on an input where the C is well-defined. All
cases with the same cause are one defect. We report defects and the translations that contain them;
case counts only show how often a defect appears. A divergence is a \emph{wrong output} (a different return value, exit code or primary output), a \emph{crash} (a Rust panic or segmentation fault where the C completes), a \emph{stderr-only difference}, or a \emph{build failure or hang}; only the first two make defects.

\begin{figure*}[t]
\centering
\includegraphics[width=\textwidth]{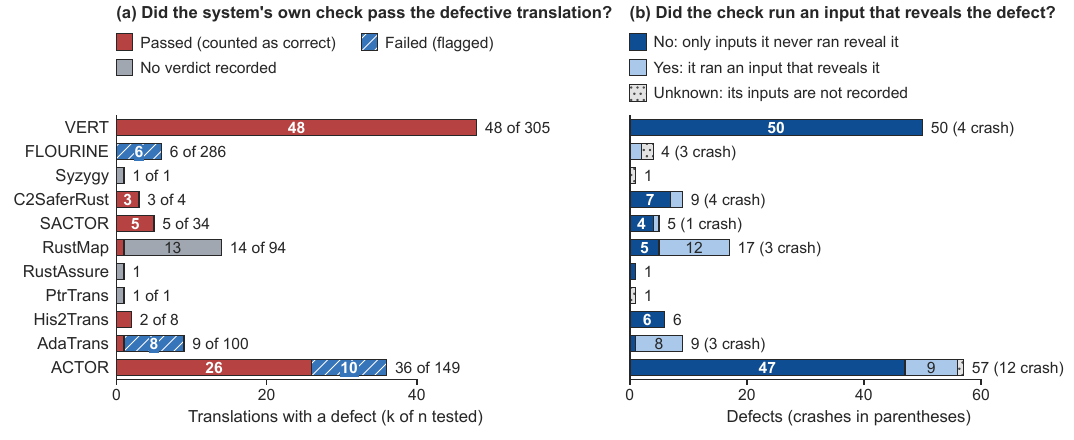}
\caption{Re-test outcomes per system. (a)~Translations with a defect, by what the system's own check
recorded: passed (\emph{accepted}), failed (\emph{rejected}), or no verdict; ``$k$ of $n$'':
defective of tested translations. (b)~Defects, crashes in parentheses, by whether the check ran an
exposing input: no (\emph{held out}), yes, or unknown (check inputs not recorded).}
\label{fig:outcomes}
\end{figure*}

\para{What we do not count.}
We do not count a divergence that occurs only (i)~where the C has undefined behavior,
(ii)~on inputs that break a precondition the original states in its name, comments or
documentation, (iii)~in a value the C standard leaves unspecified, (iv)~in a callback that frees
memory, or (v)~when the C passes a negative value to a \texttt{scanf} \texttt{\%u} or \texttt{printf} \texttt{\%x}/\texttt{\%u} conversion, which the standard does not define. We added rules (iii)-(v) during triage, after seeing the cases they cover. Of $180$ candidate defects, $12$ fall to
(i) and (ii) and $3$ to (iii)-(v); $4$ more are hangs or build failures, and $1$ has no cause we could confirm. An LLM-based assistant grouped cases into defects and assigned their classes and exclusions. An author, blind to these labels, labeled a stratified sample of $12$ cases ($9$ defects, $3$ exclusions). The two agreed on the cause for all $9$ and the class for $8$, and differed on $2$ count decisions: a precondition call, and an ACTOR parsing defect our rule (v) had dropped, which a re-run confirmed and we now count.

\para{What the systems' checks recorded.}
For each defective translation we look up what the system's own check recorded. It is \emph{accepted} if the artifact records a pass and \emph{rejected} if it records a failure; for a check with tiers, a pass at any tier the system counts as success is a pass. If the artifact records nothing, we re-run the documented check on the system's own inputs; if we cannot, the translation has \emph{no record}. An input is \emph{held out} if the system's check (in the survey: its repair loop) never
ran it. We compare against the inputs the artifact records or, for VERT, those its released commands run; otherwise held-out status is \emph{not measurable}.

\para{How we run and compare.}
We build the original C and the system's released Rust, run both on the same inputs, and compare
every observable output: a function's return value (in four systems, floats within $10^{-6}$: relative, or absolute below magnitude $1$); a
program's standard output, exit code, standard error and written files; and a library's sequence of results and callbacks. Where a system's check compares an output, we never compare it more strictly (for RustMap's Rosetta programs we
normalize whitespace, case and float precision). Where our compiler differs from the one an
artifact pins or documents (VERT, RustAssure, SACTOR, ACTOR), rebuilding with that compiler changes no verdict.
We build each system's Rust in the profile its own pipeline uses (His2Trans: debug and release; FLOURINE: release with overflow checks). We re-run every counted crash and the $58$ wrong outputs in integer code in the other profile: $86$ of $88$ still diverge from the C; the other two are FLOURINE overflow panics.

\para{Undefined behavior in the C.}
Our oracle is the C that GCC and Clang build for LP64 Linux with glibc. We take its documented
implementation-defined choices (e.g.\ arithmetic right shift, 64-bit \texttt{long}) as part of the
C. On an input where the C has UB, any Rust result is consistent with the C. A divergence therefore
counts only on an input where the C (i)~passes the address and UB sanitizers (ASan, UBSan) under
both GCC and Clang (including float-to-integer overflow); (ii)~gives identical results at
\texttt{-O0}, \texttt{-O1} and \texttt{-O2}, which can reveal uninitialized reads that ASan does not detect; and
(iii)~gives identical results with signed and unsigned \texttt{char}. Per-system scripts apply this rule; an input whose sanitizer,
optimization-level or \texttt{char} run fails or times out does not count. For VERT the full rule re-checks every exposing input. We exclude by hand two ACTOR defects: the C of
\texttt{underhanded-c-luggage} reads stale stack memory, which the sanitizers miss, and that of
\texttt{strcmp} passes an out-of-range value to \texttt{atoi}.

\para{How we generate inputs.}
Fixed scripts generate every input except three that an author wrote for one ACTOR defect; we use no language model. The scripts follow the
shape of the code under test. A function's integer arguments take the ends of their range, powers of two $\pm1$ and values in between, fewer per argument at higher arity, and its array arguments take fixed patterns. A program runs with single command-line flags and pairs of them, up to $3000$ cases per utility. It reads a fixed set of files (empty, text, binary, multibyte, NUL-delimited, large) from standard input, a file, a missing file or a directory. A library
runs fixed sequences of API calls, and a compressor a fixed set of byte strings, one of them
incompressible.

\para{How we test our harnesses.}
A harness that misses bugs would make wrong translations look correct, so we measure how many
planted bugs ours catch, by mutation analysis~\cite{mutation-analysis,mutation-survey}. We plant one bug at a time in a translation that agrees with the C: we replace a relational or arithmetic
operator, shift a constant by one, or, for VERT, compute an integer expression in \texttt{f32}. We
then re-run the harness. It \emph{catches} the planted bug, or \emph{mutant}, if it reports a wrong output or a crash on one of our inputs where the C has no UB. A difference only in standard error, a hang, or a failure of the system's own tests does not count. We report the share of compiled mutants caught per system. This share is a lower bound
where some mutants hang (VERT, AdaTrans, ACTOR) and an upper bound where the records do not separate
standard-error and timeout catches from the rest (SACTOR). 

\section{Results}
\label{sec:results}

Every system releases defective translations, and in seven of them the system's own check accepted
some (\cref{fig:outcomes}a). Most defects, $121$ of $160$, appear only on inputs the system's check
never ran (\cref{fig:outcomes}b). \Cref{fig:snip-vert,fig:snip-fluorine,fig:snip-c2saferrust,fig:snip-his2trans}
show four defects as released. All runs use Docker images
built from digest-pinned bases, with the C and the Rust built for the same architecture.

\begin{figure}[t]
\centering
\includegraphics[width=\columnwidth]{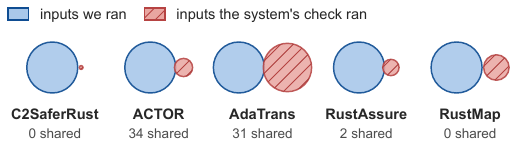}
\caption{Inputs we ran and inputs the system's own check ran, where its artifact records them;
areas and overlaps are to scale within each pair. We ran $8574$, $9130$, $8310$, $360$
and $20$ inputs; the checks ran $47$, $1164$, $7477$, $37$ and $5$, respectively (ACTOR: executables; RustMap: bzip2; RustAssure: $37$ listed, $9$ reached).}
\label{fig:overlap}
\end{figure}

\para{Overlap with the checks' inputs.}
Our inputs rarely coincide with those the systems' checks ran (\cref{fig:overlap}), which is why
most defects are held out. None of our FLOURINE inputs is among its logged examples. For VERT we take
the inputs its released commands run: none for the fuzzer, which ships no corpus for these
functions, and $n\in\{0,1\}$, which Kani draws for the Rust copy of the C only. Syzygy, PtrTrans, the ACTOR libraries and RustAssure's per-function check
record no check inputs; \cref{fig:overlap} shows RustAssure's libcsv composition test.

\subsection{VERT}
\label{sec:vert}

VERT~\cite{vert} accepts a translation that passes two checks against an \emph{rWasm copy} of the C,
the C re-expressed as Rust via WebAssembly: the fuzzing front-end \emph{bolero}~\cite{bolero} and,
where it can, the bounded model checker \emph{Kani}~\cite{kani}. We test $189$ integer functions
(arguments in $[0,512]$, $80$ to $100$ inputs each) and $116$ array and float functions. Of these, $48$ contain $50$ defects ($46$ wrong outputs, $4$ crashes), and VERT's records mark all $48$ as passing bolero and $21$ also bounded Kani (\cref{fig:snip-vert} shows one). In $8$ of them the C returns a 64-bit integer: in $2$ the Rust computes a wrong value, and in $6$ it returns a 32-bit one that overflows. The checks pass these translations because, as released, neither compares the programs on an exposing input. The Kani harness
evaluates the Rust at the fixed value $12$ before the rWasm copy draws a symbolic $n\in\{0,1\}$, so the equivalence assertion is unreachable in all $21$ defective passes. The bolero command only replays a saved corpus, which $2$ of $506$ benchmarks ship. Fuzzing each for $30$\,s instead reports a failure for $24$ of the $36$ defective scalar functions, only $6$ of them real divergences where the C has no UB. VERT's records do not say which mode produced their passes. Our
harnesses catch at least $77.7\%$ (scalar) and $59.7\%$ (array) of planted bugs. We do not count $6$ further functions that diverge only on negative or larger inputs, outside the domain we test.

\begin{figure}[t]
\centering
\input{img/snip-vert}
\caption{VERT's \texttt{HEXAGONAL\_NUMBER}: the Rust sums a series that equals $n^2$, where the C computes
$n(2n-1)$. VERT's records mark it as passing bolero and bounded Kani.}
\label{fig:snip-vert}
\end{figure}

\subsection{FLOURINE}
\label{sec:flourine}

FLOURINE~\cite{flourine} accepts a translation that its differential fuzzer finds equivalent. Of the
$286$ shipped \texttt{success} files we test ($101$ scalar functions, tested on up to $2000$ signed
inputs each with $|x|\le2^{20}$, and $185$ struct accessors), $6$ contain one of $4$ defects. A byte packing became a sum because the Rust shifts a \texttt{u8} (\cref{fig:snip-fluorine}). In two functions a mask \texttt{(1<{}<len)-1} became \texttt{(1>{}>len)-1}, which underflows, and in a third a multiplication overflows where the C's unsigned arithmetic wraps; all three panic in FLOURINE's build. FLOURINE's own logs reject all six: each is an experiment's first candidate, logged as not equivalent; FLOURINE does not ship the accepted one.
This mismatch affects $20$ of the $286$ files. We do not count divergences on ranges whose minimum
exceeds the maximum, though FLOURINE's fuzzer does. Our scalar harness catches $88\%$ of $632$ planted bugs.

\begin{figure}[t]
\centering
\input{img/snip-fluorine}
\caption{FLOURINE's byte packing: shifting a \texttt{u8} by $8$, $16$ or $24$ wraps to a no-op, so the Rust adds
the bytes. FLOURINE's own log rejects this file, which its artifact ships as a success.}
\label{fig:snip-fluorine}
\end{figure}

\subsection{RustAssure}
RustAssure~\cite{rustassure} checks each translated function by differential symbolic execution
under KLEE~\cite{klee}. Of the $151$ GPT-4o functions in its archived run, $2$ run in isolation (both agree), so we also test the two libraries it ships in runnable form. In libcsv with \texttt{CSV\_EMPTY\_IS\_NULL} set, the C passes the field callback a NULL
pointer for an empty unquoted field. GPT-4o's \texttt{csv\_parse} instead made the parameter a
non-nullable slice, so \texttt{a,{},b} and \texttt{a,"",b} become indistinguishable ($12$ of $360$
cases). RustAssure records no verdict for \texttt{csv\_parse}. In its artifact, the per-function check also
replaces calls through function pointers with stubs, so it cannot observe a callback. In u8c, a decoding divergence that RustAssure's own bug list records depends on
the signedness of \texttt{char} in the C, so we do not count it.

\subsection{C2SaferRust}
\label{sec:c2saferrust}

C2SaferRust~\cite{c2saferrust} rewrites C2Rust's rule-based translation of
coreutils~\cite{coreutils} with an LLM. Its artifact keeps a rewrite only if the
test scripts that passed on the starting translation still pass. We test \texttt{cat}, \texttt{head}, \texttt{tail} and \texttt{uniq}
(four of its seven utilities) on $8574$ cases ($5942$ distinct), including the file \texttt{uniq}
writes when given an output operand. We find $9$ defects in three, all from rewrites the log records as passing (the C2Rust starting point agrees with the C). \texttt{tail} binds its parsed \texttt{-c}/\texttt{+N} mode to locals and never stores them, so \texttt{-c N} runs as
\texttt{-n N} (\cref{fig:snip-c2saferrust}). \texttt{head} drops a final partial buffer when it reads a pipe, and on x86-64 its inherited C2Rust code reads regular files that way too. It loses $4{,}095$ bytes of its own acceptance test's file; the test misses this because it sends the output to \texttt{/dev/full}. Four defects
crash. In our re-run, the shipped translation fails $2$ of the $27$ tests its loop selects across its seven utilities, one because of
the \texttt{tail} defect. \texttt{uniq} also segfaults, but the C performs \texttt{NULL+0} pointer
arithmetic (UB in C17) on every input \texttt{uniq} reads, so we do not count it. Our harness catches $20\%$ of $1892$ bugs we plant in the C2Rust starting point ($19.2\%$ of $1074$ in C2SaferRust's output). Of the $935$ survivors we could not show equivalent, $619$ lie in code our inputs never execute.

\begin{figure}[t]
\centering
\input{img/snip-c2saferrust}
\caption{C2SaferRust's \texttt{tail}: the Rust parses \texttt{-c} into a new local instead of the global
\texttt{count\_lines}, so \texttt{tail -c 3} prints three lines, not three bytes. The log records this
rewrite as passing.}
\label{fig:snip-c2saferrust}
\end{figure}

\subsection{Syzygy}
\label{sec:others}

Syzygy's~\cite{syzygy} translated Zopfli compressor emits nothing on empty input, where
the C's \texttt{do-while} always emits a final block. Its paper reports that it
passed $10^6$ random inputs of $10$ to $10^7$ characters, a range that excludes the empty input.
It ships neither a verdict nor those inputs.

\subsection{RustMap}
RustMap~\cite{rustmap} ships a translated bzip2 and $124$ Rosetta Code programs, of which $93$
build in both languages; with bzip2 we test $94$. Its bzip2 check salvages the Rust's output with
\texttt{bzip2recover}, decompresses it with the C, and passes on RustMap's test files. The translation
has four defects. Input bytes $\ge$\texttt{0x80} become \texttt{0xFF}, the Rust encodes runs of identical bytes wrongly, and a 1-byte input panics. Every output also starts with $5000$ zero bytes, which
plain \texttt{bzip2~-d} rejects, also on RustMap's own files. The Rosetta programs read no input; RustMap's own run never reaches $2$ of them.
Its paper reports that all $125$ give the same output as the C. Its shipped whole-log diff,
however, differs on every program, so it records no per-program verdict. After normalizing, the outputs differ for the $13$ programs we count (e.g.\ a Roman-numeral decoder reads
\texttt{MCmxC} as $1000$; two knapsack solvers keep $1$ of $22$ items and $1$ of $9$). Six more implement a different task, which we set aside. We run these on x86-64.

\subsection{His2Trans}
\label{sec:new}

His2Trans~\cite{his2trans} reports how many of a translated library's tests pass. We test the $8$
libraries of its open-source benchmark function by function and find $6$ defects in $2$. Its
results and our re-run of its tests pass all $8$. \texttt{buffer} has five
(e.g.\ \texttt{buffer\_append\_n} ignores its length argument, \cref{fig:snip-his2trans}) and \texttt{rgba} one.
For \texttt{buffer}, His2Trans runs $14$ tests of its own instead of the project's, none of which calls \texttt{buffer\_append\_n}; the project's own test for it fails on the shipped Rust. One
\texttt{buffer} defect makes the Rust correct where the C His2Trans ships is not; it counts because the C is
the oracle. We do not count \texttt{genann}'s replaced \texttt{rand()}, whose sequence ISO C leaves
unspecified. Our harness catches $53\%$ of $113$ planted bugs.

\begin{figure}[t]
\centering
\input{img/snip-his2trans}
\caption{His2Trans's \texttt{buffer\_append\_n} ignores its length argument, so appending one byte of
\texttt{"hello"} appends all five. His2Trans's results and our re-run of its tests pass this library.}
\label{fig:snip-his2trans}
\end{figure}

\subsection{SACTOR}
SACTOR~\cite{sactor} accepts a translation if the C project's end-to-end tests pass against
the Rust through the C interface. SACTOR does not ship its headline translations; we test
the $34$ shipped ones that are complete, keep the C interface (its unidiomatic stage) and build. Five contain a defect, and SACTOR's check
accepts all five. In Holdem-Odds, a nested stub returns \texttt{true} for every card; SACTOR's own
test prints the failure but exits $0$, and the check reads only the exit code. Two other
defects corrupt a buffer or crash at the C interface. Four of the five diverge only on inputs SACTOR's tests do
not run. Our harnesses catch at most $63\%$ of $109$ planted bugs.

\subsection{PtrTrans}
PtrTrans~\cite{ptrtrans} ships its own Rust for one of its $16$ projects, \texttt{quadtree},
whose search never descends below the root. Its paper reports \texttt{quadtree} as $87.5\%$
equivalent; it ships no per-function verdict or test inputs. Our harness catches $91\%$ of $64$ bugs we plant in the crate after repairing its search.

\subsection{ACTOR}
ACTOR~\cite{actoruw} translates a DARPA test corpus. It counts a translation as correct if it passes
the corpus's hidden test vectors. We test the $149$ of its $211$ translations (all $61$ executables
and $88$ libraries) whose interfaces take only scalars or strings and that have a linkable C oracle.
Of these, $36$ contain $57$ defects; ACTOR counted $26$ of the $36$ correct. About half the defects read
numbers differently: Rust's \texttt{str::parse} rejects inputs that \texttt{scanf} or \texttt{strtol}
accept, such as \texttt{" 12"} or \texttt{3.14} for an integer.  Of the $26$, $14$ differ only in parsing
numbers, $9$ crash on non-UTF-8 input or missing tokens, and $3$ differ in NaN spelling, long lines,
or a right shift of a negative value. Our harness, run on ACTOR's $1129$ replayable recorded inputs, finds defects only in translations ACTOR rejected. Our harness catches at least $42.2\%$ of $393$ bugs we plant in $84$ of its $94$ agreeing translations.

\subsection{AdaTrans}
AdaTrans~\cite{adatrans} counts a translated LeetCode-style program as solved if it passes its
repair loop's example tests and then a fuzzing oracle. The loop compares the Rust with per-problem expected outputs. For $7$ of the $100$ problems whose translation builds, the C does not
reproduce these outputs. Of the $100$, $9$ contain a defect, and AdaTrans counts one of them as solved, though it panics at $n=1$, a value no fuzzing input reaches. Three of the nine follow the recorded outputs, not the C. Our harness catches at least $75\%$ of $410$ planted bugs.

\section{Discussion}
\label{sec:discussion}

\begin{finding}{Findings}
\textbf{RQ1.} \emph{Do translations preserve the C's behavior?} Not always. All eleven systems release defective translations: $160$ defects in $126$ of the translations we test. Of the defects, $93$ are outright errors. Another $54$ are \emph{edge-input} defects, on malformed or extreme inputs on which
the C is defined; in $12$ the Rust follows a bug fix or a
recorded expected output instead of the C, and $1$ right-shifts a negative value.\par\smallskip
\textbf{RQ2.} \emph{What did the checks record, and did they run an exposing input?} Their own checks accepted $86$ of these translations, in seven systems; $56$ contain an outright error. For $81$ of the $86$, the released check never compared the programs on an exposing input. For three
of the other five, it ran one but compared only part of the result: the exit code of a test suite that exits $0$ even when its tests fail, the exit status and error message only, or an output that a recovery tool repaired.\par\smallskip
\textbf{RQ3.} \emph{How robust are these findings?} Every counted defect has an exposing input where the C passes our UB rule, and we confirmed each in the source. All but two of the $30$ crashes and $58$ wrong outputs in integer code also diverge in the other build profile. Where our compiler differs from the one an artifact pins, rebuilding with that compiler changes no verdict. Our harnesses catch from $20\%$ of planted bugs (C2SaferRust) to $91\%$ (PtrTrans).
\end{finding}

\para{What the checks miss.}
Where artifacts record inputs, $105$ of the $109$ defects in accepted translations lie on inputs the
check never ran. Of the other four, three pass the partial comparisons in RQ2 and one (C2SaferRust's \texttt{tail}) fails our re-run of the check.
Where a check compared fully, it worked: ACTOR's inputs expose defects only in translations it
rejected, and FLOURINE's fuzzer rejected every defective file we found.

\para{Recommendations.}
Four practices address the problems we report: comparing every output on held-out inputs,
checking that a verifier reaches its assertion, releasing exactly what the check accepted, and reporting tests of the check.

\section{Related work}
\label{sec:related}

We use differential testing~\cite{mckeeman,csmith} with the original program as the
oracle~\cite{oracle-problem}. Sanitizers~\cite{asan} and mutation
analysis~\cite{mutation-analysis,mutation-survey} validate our oracle and harnesses. None of VERT's $90$ recorded
bounded-Kani passes reaches the assertion; $81$ are \emph{vacuous}~\cite{vacuity-beer,vacuity-kv}. EvalPlus~\cite{evalplus} showed that weak tests overstate the
correctness of LLM-generated code. Pan et al.~\cite{lost-in-translation} catalogued the bugs LLMs
introduce when translating code. Among
C-to-Rust systems, ACToR~\cite{actor} (not the ACTOR we test) searches adversarially for inputs
where C and Rust disagree, and Heimdall~\cite{heimdall} tests its symbolic check on
known-wrong translations.

\section{Availability}
\label{sec:repro}

The artifact, with all files and results, is at \url{https://github.com/muhammad-hassnain/retest-c-to-rust}.

\para{Use of AI.} Claude Code (Claude Opus 5.5) helped write code, triage divergences, label defects and draft this paper.

\section{Conclusion}
\label{sec:conclusion}

We re-tested eleven C-to-Rust systems against their original C: all release defective translations, and their checks accepted $86$ of them. A success rate earns trust when authors report how they tested the check and release exactly what it accepted.

\bibliographystyle{IEEEtran}
\bibliography{ref}

\end{document}